\documentclass[%
 showpacs,
 amsmath,amssymb,
]{revtex4-1}

\usepackage{graphicx}
\usepackage{dcolumn}
\usepackage{bm}

\usepackage{CJK}

\usepackage{color}
\usepackage{amssymb}
\usepackage{amsfonts}

\usepackage[colorlinks,urlcolor=blue,linkcolor=blue,citecolor=blue,anchorcolor=blue]{hyperref}

\begin{document}

\preprint{APS/123-QED}

\title{New face-centered photonic square lattices with flat bands}

\author{Yiqi Zhang$^1$}
\email{zhangyiqi@mail.xjtu.edu.cn}
\author{Milivoj R. Beli\'c$^{2}$}
\author{Changbiao Li$^1$}
\author{Zhaoyang Zhang$^1$}
\author{Yanpeng Zhang$^{1}$}
\author{Min Xiao$^{3,4}$}
\affiliation{%
 $^1$Key Laboratory for Physical Electronics and Devices of the Ministry of Education \& Shaanxi Key Lab of Information Photonic Technique,
Xi'an Jiaotong University, Xi'an 710049, China \\
$^2$Science Program, Texas A\&M University at Qatar, P.O. Box 23874 Doha, Qatar \\
$^3$Department of Physics, University of Arkansas, Fayetteville, Arkansas, 72701, USA\\
$^4$National Laboratory of Solid State Microstructures and School of Physics, Nanjing University, Nanjing 210093, China
}%

\date{\today}

\begin{abstract}
\noindent
  We report two new classes of face-centered photonic square lattices with flat bands which we call the Lieb-I and the Lieb-II lattices.
  There are 5 and 7 sites in the corresponding unit cells of the simplest Lieb-I and Lieb-II lattices, respectively.
  The number of flat bands $m$ in the new Lieb lattices is related to the number of sites $N$ in the unit cell by $m=(N-1)/2$.
  Physical properties of the lattices with even and odd number of flat bands are different.
  We also consider localization of light in such Lieb lattices.
  If the input beam excites the flat-band mode, it will not diffract during propagation,
  owing to the strong localization in the flat-band mode.
  For the Lieb-II lattice, we also find that the beam will oscillate and still not diffract during propagation,
  because of the intrinsic oscillating properties of certain flat-band modes.
  The period of oscillation is determined by the energy difference between the two flat bands.
  This study provides a new platform for the investigation of flat-band modes.
\end{abstract}

\maketitle

\section{Introduction}

A flat band --- the bandwidth of which is zero --- plays an essential role in studying strongly correlated phenomena,
because the eigenstates of the flat band are highly degenerate \cite{sun.prl.106.236803.2011}.
By the same token, the flat band can significantly help in realizing nondiffracting and localized states,
because both the first-order and the second-order derivatives of the band are zero.
Indeed, the localized flat-band modes \cite{vicencio.prl.114.245503.2015,mukherjee.prl.114.245504.2015}
and image transmission \cite{xia.ol.41.1435.2016,vicencio.jo.16.015706.2014,zong.oe.24.8877.2016} based on the flat-band modes have been reported
in the Lieb lattice \cite{apaja.pra.82.041402.2010,shen.prb.81.041410.2010,leykam.pra.86.031805.2012,nita.prb.87.125428.2013,guzman-silva.njp.16.063061.2014,zhang.rrp.68.230.2016}
and the kagome lattice \cite{atwood.nm.1.91.2002,boguslawski.apl.98.061111.2011,jo.prl.108.045305.2012,nakata.prb.85.205128.2012}
that possess one flat band under the tight-binding approximation, when only the nearest-neighbor (NN) hopping is considered.
Up to date, there is a variety of models reported that exhibit flat bands \cite{aoki.prb.54.R17296.1996,deng.jssc.176.412.2003,wu.prl.99.070401.2007,tasaki.epjb.64.365.2008,crespi.njp.15.013012.2013,jacqmin.prl.112.116402.2014,zhang.lpr.9.331.2015,baboux.prl.116.066402.2016},
among which the Lieb lattice is perhaps the simplest one.
The Lieb lattice, which is not a Bravais lattice, has 3 sites in the unit cell and represents a kind of face-centered square lattice.
If one can view the Lieb lattice as originating from a square lattice, a natural question arises:
can one artificially fabricate other kinds of face-centered square lattices?
For example, a face-centered square lattice with 5 or 7 sites in the unit cell.
Since there are 3 sites in the unit cell of a usual Lieb lattice,
the face-centered square lattices with 5 and 7 sites introduced here represent novel members of the face-centered square lattice family,
which we call the Lieb-I and the Lieb-II lattices.
We believe such new face-centered lattices can be realized by the direct laser writing technique \cite{szameit.jpb.43.163001.2010},
which is adopted in recent research on topological photonics \cite{rechtsman.nature.496.196.2013,rechtsman.prl.111.103901.2013,plotnik.nm.13.57.2014}.

In this paper, we first investigate the dispersion relations (the band structure) of such face-centered photonic square lattices,
based on the tight-binding method.
We find that such photonic square lattices possess more than one flat band,
which is quite different from  other flat-band models reported in previous literature,
where usually only one flat band is observed.
The number of flat bands $m$ is associated with the number of sites $N$ included in the unit cell, i.e., $m=(N-1)/2$.
Thus, for $N$ odd, one can have an even or an odd number of flat bands.
In addition to many potential applications in image processing, telecommunication and sensing, and
strongly correlated states, we believe that the study reported in this paper provides a new platform for investigating the flat bands.

The organization of the paper is as follows.
In Secs. \ref{case4} and \ref{case5} we investigate the dispersion relation as well as the localization of light due to flat-band modes of the Lieb-I and Lieb-II lattices, respectively;
in Sec. \ref{conclusion}, we conclude the paper.

\section{Lieb-I lattice}
\label{case4}

\begin{figure*}[htbp]
\centering
  \includegraphics[width=\textwidth]{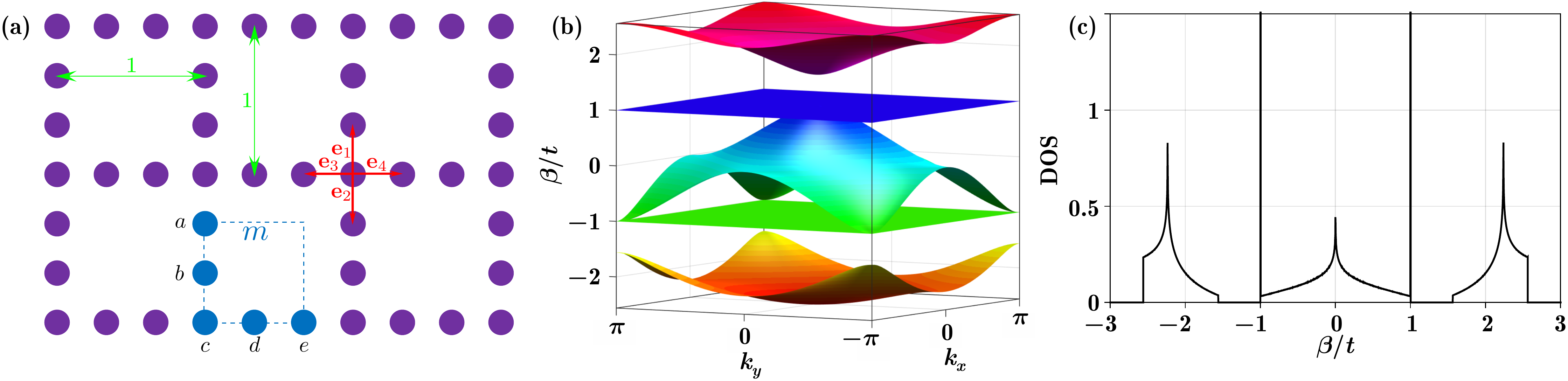}
  \caption{(Color online)
  (a) Edge-centered photonic square lattice with 5 sites in the unit cell,
  which are labeled as $a$, $b$, $c$, $d$, and $e$.
  The letter $m\in\mathbb{Z}$ represents the index of the unit cell.
  The hopping unit vectors among NN sites are labeled by $\mathbf{e}_1$, $\mathbf{e}_2$, $\mathbf{e}_3$, and $\mathbf{e}_4$.
  (b) Dispersion relation. From top to bottom, the bands are $\beta_{1\sim5}$, respectively.
  (c) Density of states per unit cell as a function of the eigenvalues (in units of $t$).}
  \label{fig1}
\end{figure*}

The simplest novel Lieb-I lattice is displayed in Fig. \ref{fig1}(a), in which the lattice sites enclosed in the dashed square form a unit cell and the period of the lattice is set to be 1.
We assume that the hopping among lattice points only happens between the NN sites, with $t$ being the hopping strength.
The propagation dynamics of light in this discrete model can be described by the discrete coupled Schr\"odinger equations \cite{lederer.pr.463.1.2008,longhi.lpr.3.243.2009,kartashov.rmp.83.247.2011}
\begin{subequations}\label{eq1}
\begin{equation}\label{eq1a}
i\frac{\partial a_m}{\partial z} = t \sum_{\boldsymbol{R}_m,\mathbf{e}_i} (b_m + c_m),
\end{equation}
\begin{equation}\label{eq1b}
i\frac{\partial b_m}{\partial z} = t \sum_{\boldsymbol{R}_m,\mathbf{e}_i} (a_m + c_m),
\end{equation}
\begin{equation}\label{eq1c}
i\frac{\partial c_m}{\partial z} = t \sum_{\boldsymbol{R}_m,\mathbf{e}_i} (a_m + b_m + d_m + e_m),
\end{equation}
\begin{equation}\label{eq1d}
i\frac{\partial d_m}{\partial z} = t \sum_{\boldsymbol{R}_m,\mathbf{e}_i} (c_m + e_m),
\end{equation}
\begin{equation}\label{eq1e}
i\frac{\partial e_m}{\partial z} = t \sum_{\boldsymbol{R}_m,\mathbf{e}_i} (c_m + d_m),
\end{equation}
\end{subequations}
where $\boldsymbol{R}_m$ is the position of the $m$th unit cell,
and ${\bf e}_i$ are the 4 vectors connecting the neighboring sites,
as displayed in Fig. \ref{fig1}(a).
It is assumed that an array of optical waveguides is arranged in the form of Lieb-I lattice and that light propagates perpendicular to the lattice, along waveguides.
We are looking for the solutions to Eq. (\ref{eq1}) of the form \cite{szameit.pra.84.021806.2011}:
\begin{align*}
a_m = & a_\mathbf{k} \exp[i(\beta z + \boldsymbol{R}_m \cdot \mathbf{k})], ~\\
b_m = & b_\mathbf{k} \exp[i(\beta z + \boldsymbol{R}_m \cdot \mathbf{k})], ~\\
c_m = & c_\mathbf{k} \exp[i(\beta z + \boldsymbol{R}_m \cdot \mathbf{k})], ~\\
d_m = & d_\mathbf{k} \exp[i(\beta z + \boldsymbol{R}_m \cdot \mathbf{k})], ~\\
e_m = & e_\mathbf{k} \exp[i(\beta z + \boldsymbol{R}_m \cdot \mathbf{k})],
\end{align*}
in which case one can rewrite Eq. (\ref{eq1}) in the matrix form, as an eigenvalue problem
\begin{align}\label{eq2}
  {H}_{\rm TB} |\beta,\mathbf{k}\rangle  =  \beta |\beta,\mathbf{k}\rangle
\end{align}
with
$|\beta,\mathbf{k}\rangle =
\begin{bmatrix}
    a_{\bf k}, & b_{\bf k}, & c_{\bf k}, & d_{\bf k}, & e_{\bf k}
\end{bmatrix}^T$
and $H_{\rm TB}$, the tight-binding Hamiltonian of the system, being
\begin{equation}\label{eq3}
H_{\rm TB} = -t
\begin{bmatrix}
  0             & \exp(-ik_y/3)   & \exp(ik_y/3)   & 0             & 0             \\
  \exp(ik_y/3)  & 0               & \exp(-ik_y/3)  & 0             & 0             \\
  \exp(-ik_y/3) & \exp(ik_y/3)    & 0              & \exp(ik_x/3)  & \exp(-ik_x/3) \\
  0             & 0               & \exp(-ik_x/3)  & 0             & \exp(ik_x/3)  \\
  0             & 0               & \exp(ik_x/3)   & \exp(-ik_x/3) & 0
\end{bmatrix}.
\end{equation}
Clearly, the matrix in Eq. (\ref{eq3}) is equal to its own conjugate transpose,
thus it is a Hermitian matrix, and the corresponding eigenvalues are completely real.
Diagonalizing $H_{\rm TB}$, one obtains the eigenvalues $\beta_{1\sim5}$, as
\begin{subequations}\label{eq4}
\begin{align}\label{eq4a}
\beta_1 =  \sqrt[3]{-\frac{q}{2}+\sqrt{\left(\frac{q}{2}\right)^2+ \left(\frac{p}{3}\right)^3}} + 
 \sqrt[3]{-\frac{q}{2}-\sqrt{\left(\frac{q}{2}\right)^2+ \left(\frac{p}{3}\right)^3}},
\end{align}
\begin{align}\label{eq4b}
  \beta_2 =t,
\end{align}
\begin{align}\label{eq4c}
\beta_3 =  \frac{-1 - \sqrt{3}i}{2} \sqrt[3]{-\frac{q}{2}+\sqrt{\left(\frac{q}{2}\right)^2+ \left(\frac{p}{3}\right)^3}} + 
 \frac{-1 + \sqrt{3}i}{2} \sqrt[3]{-\frac{q}{2}-\sqrt{\left(\frac{q}{2}\right)^2+ \left(\frac{p}{3}\right)^3}},
\end{align}
\begin{align}\label{eq4d}
\beta_4 =-t,
\end{align}
and
\begin{align}\label{eq4e}
\beta_5 =  \frac{-1+ \sqrt{3}i}{2} \sqrt[3]{-\frac{q}{2}+\sqrt{\left(\frac{q}{2}\right)^2+ \left(\frac{p}{3}\right)^3}} + 
\frac{-1- \sqrt{3}i}{2} \sqrt[3]{-\frac{q}{2}-\sqrt{\left(\frac{q}{2}\right)^2+ \left(\frac{p}{3}\right)^3}},
\end{align}
\end{subequations}
with $p=-5t^2$ and $q=2t^3[\cos(k_x)+\cos(k_y)]$.
The eigenvalues in Eqs. (\ref{eq4b}) and (\ref{eq4d}) are independent on $k_x$ or $k_y$,
so they form the flat bands in the first Brillouin zone.
The eigenstates corresponding to the flat bands are
\begin{align}\label{eq5}
  |\beta_2,\mathbf{k}\rangle =
  \left[
  \begin{array}{l}
  -\displaystyle{\frac{\exp(ik_y/3)}{-1+\exp(ik_y)}}\\
  \\
  -\displaystyle{\frac{\exp(-ik_y/3)}{-1+\exp(-ik_y)}}\\
  \\
  0\\
  \\
  \displaystyle{\frac{\exp(-ik_x/3)}{-1+\exp(-ik_x)}}\\
  \\
  \displaystyle{\frac{\exp(ik_x/3)}{-1+\exp(ik_x)}}
  \end{array}
  \right],~
  |\beta_4,\mathbf{k}\rangle =
  \left[
  \begin{array}{l}
  -\displaystyle{\frac{\exp(ik_y/3)}{1+\exp(ik_y)}}\\
  \\
  -\displaystyle{\frac{\exp(-ik_y/3)}{1+\exp(-ik_y)}}\\
  \\
  0\\
  \\
  \displaystyle{\frac{\exp(-i k_x/3)}{1+\exp(-ik_x)}}\\
  \\
  \displaystyle{\frac{\exp(i k_x/3)}{1+\exp(ik_x)}}
  \end{array}
  \right],
\end{align}
from which one can find that the amplitudes on sites $a$ and $b$ (as well as $d$ and $e$) are complex conjugate.

\begin{figure}[htbp]
\centering
  \includegraphics[width=0.5\columnwidth]{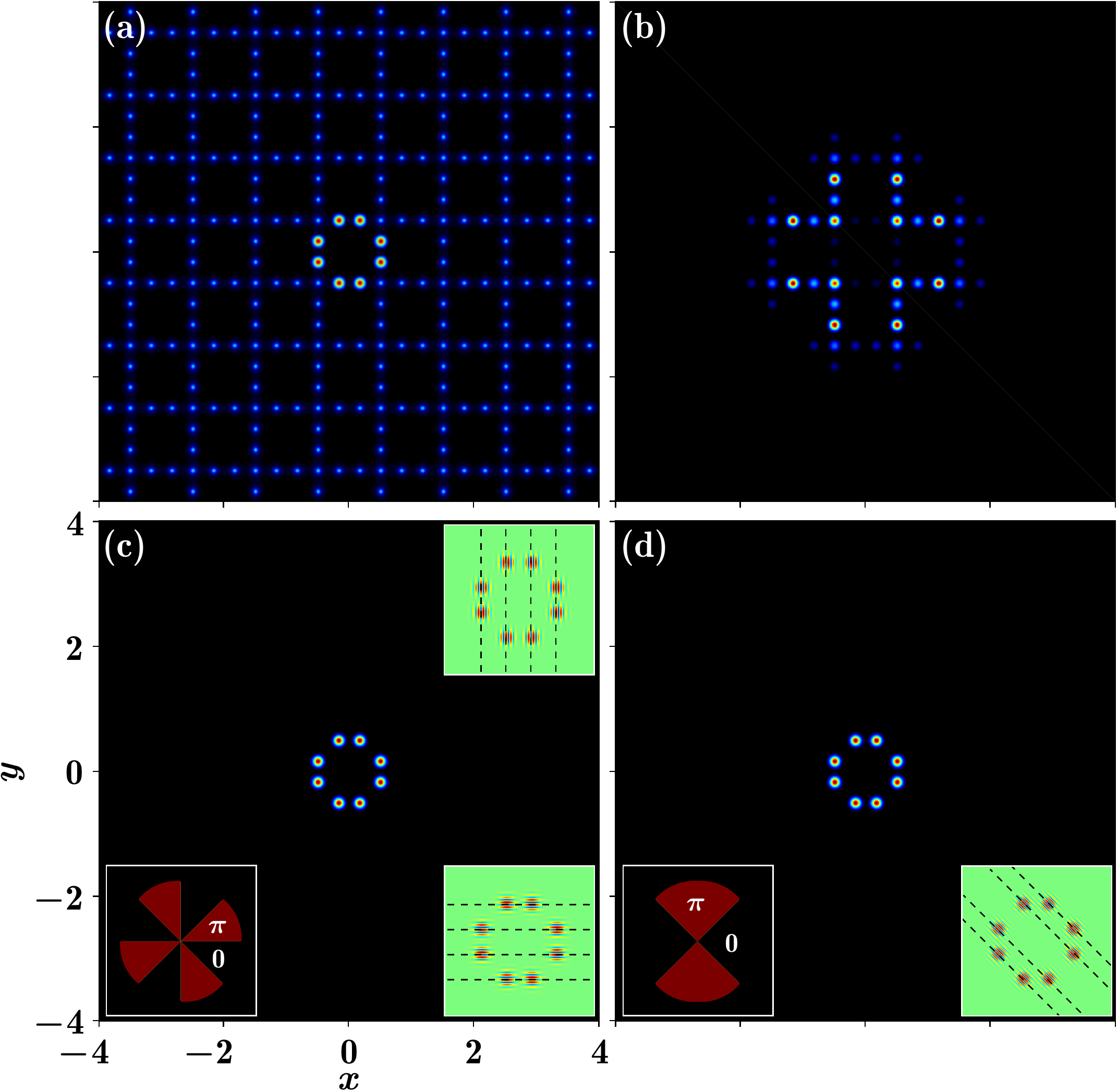}
  \caption{(Color online)
  (a) Intensity of the input that contains 8 peaks, which are exactly located at the sites, except the ones at the corners.
  (b) Discrete diffraction of the input if the peaks are in-phase.
  (c) Output intensity of the beam when the peaks are out-of-phase. Left-bottom inset: the phase of the input.
  Right top and bottom insets: interferograms of the output beam with tilted plane waves.
  (d) Figure setup is as (c), but for the input with a different condition.
  }
  \label{fig2}
\end{figure}

In Fig. \ref{fig1}(b), we show the numerical dispersion relation of the Lieb-I lattice in the first Brillouin zone.
One finds that, different from the usual Lieb lattice \cite{leykam.pra.86.031805.2012,nita.prb.87.125428.2013,guzman-silva.njp.16.063061.2014},
there is no particle-hole symmetry for the states $|\beta_{1\sim5},\mathbf{k}\rangle$,
even though there are two flat bands.
As an important quantity, the density of states (DOS) of a system is used to check how many states will be occupied per energy interval of a certain energy level \cite{neto.rmp.81.109.2009}.
In Fig. \ref{fig1}(c), we display the DOS of the face-centered photonic square lattice that corresponds to the band structure in Fig. \ref{fig1}(b).
One finds that in the DOS diagram, the number of states increases sharply to a very large value from zero when the energy passes $\beta=\pm t$.
The reason is clear --- there are two band gaps which lead to zero DOS,
but for a flat band, the states are numerous and degenerate, so the DOS increases sharply.

Now, we check the localization of light in the Lieb-I lattice, due to flat-band modes.
The propagation of light in such a lattice, based on Eqs. (\ref{eq1}), is presented in Fig. \ref{fig2}.
We assume that the input beam is an octupole that has eight peaks focused on sites $a$, $b$, $d$ and $e$, as shown in Fig. \ref{fig2}(a).
If the eight peaks are in-phase, then the beam undergoes discrete diffraction during propagation,
as shown in Fig. \ref{fig2}(b), which presents the output intensity distribution of the beam.

On the other hand, if the eight peaks are out-of-phase, the fundamental flat-band mode will be excited,
and the beam will remain localized and invariant during propagation.
In Fig. \ref{fig2}(c), we show the output intensity distribution of the out-of-phase input with the corresponding input phase displayed in the left-bottom inset.
As expected, the flat-band mode is excited, and the output beam intensity is the same as the input.
In order to check the phase of each peak, we also display the interferogram of the output beam,
by interfering the output beam with two tilted plane waves, as shown by the right two insets in Fig. \ref{fig2}(c).
According to the interference stripes along the dashed lines,
one finds that the initial out-of-phase structure is well preserved.
If we change the phase distribution of the eight peaks of the input [the left-bottom inset in Fig. \ref{fig2}(d)],
we find that the beam can still be well localized during propagation with the phase structure preserved,
as shown in Fig. \ref{fig2}(d) and the right-bottom inset in this panel.
From Figs. \ref{fig2}(c) and \ref{fig2}(d), one can advance the idea that both inputs with out-of-phase conditions
can excite the flat-band modes efficiently, and result in the localization of light propagating through the lattice.

Light propagating in the Lieb-II lattice offers a different behavior.

\section{Lieb-II lattice}
\label{case5}

\begin{figure*}[htbp]
\centering
  \includegraphics[width=\textwidth]{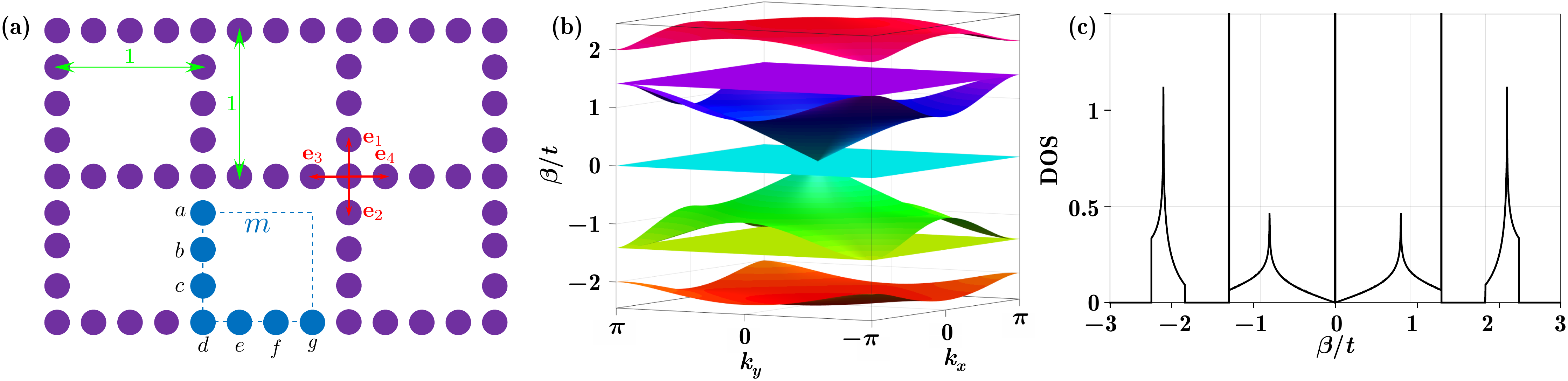}
  \caption{(Color online)
  Same as Fig. \ref{fig1}, but for the Lieb-II lattice.}
  \label{fig3}
\end{figure*}

The Lieb-II lattice is displayed in Fig. \ref{fig3}(a), and the period is still set to 1.
Following the same method as for the Lieb-I lattice in Sec. \ref{case4},
the Hamiltonian under the NN hopping approximation for the Lieb-II lattice can be written as
\begin{equation}\label{eq6}
H_{\rm TB}=-t
\begin{bmatrix}
 0 & \exp (-i k_y/4) & 0 & \exp (i k_y/4) & 0 & 0 & 0 \\
 \exp (i k_y/4) & 0 & \exp (-i k_y/4) & 0 & 0 & 0 & 0 \\
 0 & \exp (i k_y/4) & 0 & \exp (-i k_y/4) & 0 & 0 & 0 \\
 \exp (-i k_y/4) & 0 & \exp (i k_y/4) & 0 & \exp (i k_x/4) & 0 & \exp (-i k_x/4) \\
 0 & 0 & 0 & \exp (-i k_x/4) & 0 & \exp (i k_x/4) & 0 \\
 0 & 0 & 0 & 0 & \exp (-i k_x/4) & 0 & \exp (i k_x/4) \\
 0 & 0 & 0 & \exp (i k_x/4) & 0 & \exp (-i k_x/4) & 0
\end{bmatrix},
\end{equation}
the eigenvalues of which are
\begin{subequations}\label{eq7}
\begin{align}\label{eq7a}
\beta_1 = \sqrt{3 + \sqrt{5 + 2\cos(k_x) + 2\cos(k_y)}}t,
\end{align}
\begin{align}\label{eq7b}
  \beta_2 = \sqrt{2} t,
\end{align}
\begin{align}\label{eq7c}
\beta_3 = \sqrt{3 - \sqrt{5 + 2\cos(k_x) + 2\cos(k_y)}}t,
\end{align}
\begin{align}\label{eq7d}
\beta_4 = 0,
\end{align}
\begin{align}\label{eq7e}
\beta_5 = -\sqrt{3 - \sqrt{5 + 2\cos(k_x) + 2\cos(k_y)}}t,
\end{align}
\begin{align}\label{eq7f}
\beta_6 = -\sqrt{2}t,
\end{align}
and
\begin{align}\label{eq7g}
\beta_7 = -\sqrt{3 + \sqrt{5 + 2\cos(k_x) + 2\cos(k_y)}}t.
\end{align}
\end{subequations}
Clearly, the Lieb-II lattice possesses three flat bands $\beta_{2,4,6}$,
as displayed in Fig. \ref{fig3}(b).
Generally, one can obtain $m$ flat bands, if there are $2m+1$ sites in the unit cell for such novel kinds of face-centered square Lieb lattices.
From the band structure, one finds that for each state $|\beta,\mathbf{k}\rangle$,
there is a corresponding state $|-\beta,\mathbf{k}\rangle$.
That is, there exists the particle-hole symmetry,
which also exists in the usual Lieb lattice with only one flat band \cite{leykam.pra.86.031805.2012}.
Thus, one should distinguish lattices with an even number of flat bands (such as Lieb-I lattice)
from the lattices with an odd number of flat bands (such as Lieb-II lattice).

Another phenomenon that is similar to that in the usual Lieb lattice, is the existence of a single Dirac cone in the Lieb-II lattice,
which is intersected by the flat band $\beta_4$ and terminates at the other two flat bands $\beta_{2,6}$.
It is worth mentioning that the Dirac cone is located at the center of the first Brillouin zone,
which is different from the case of the usual Lieb lattice, where it is located at the corners of the first Brillouin zone.
Corresponding to the flat bands, the eigenstates are
\begin{align}\label{eq8}
  |\beta_{2,6},\mathbf{k}\rangle =  \left[
  \begin{array}{l}
  -\displaystyle{\frac{\exp(ik_y/4)}{1+\exp(ik_y)}} \\
  \\
  \pm \sqrt{2} \displaystyle{\frac{\exp(ik_y/2)}{1+\exp(ik_y)}} \\
  \\{}
  - \displaystyle{\frac{\exp(-ik_y/4)}{1+\exp(-ik_y)}} \\
  \\
  0\\
  \\
  \displaystyle{\frac{\exp(-ik_x/4)}{1+\exp(-ik_x)}} \\
  \\
  \mp \sqrt{2} \displaystyle{\frac{\exp(ik_x/2)}{1+\exp(ik_x)}} \\
  \\
  \displaystyle{\frac{\exp(ik_x/4)}{1+\exp(ik_x)}}
  \end{array}
  \right],~
  |\beta_4,\mathbf{k}\rangle = &
  \left[
  \begin{array}{l}
  -\displaystyle{\frac{\exp(ik_y/4)}{-1+\exp(ik_y)}} \\
  \\
  0 \\
  \\{}
  -\displaystyle{\frac{\exp(-ik_y/4)}{-1+\exp(-ik_y)}}  \\
  \\
  0\\
  \\
  \displaystyle{\frac{\exp(-ik_x/4)}{-1+\exp(-i k_x)}} \\
  \\
  0\\
  \\
  \displaystyle{\frac{\exp(ik_x/4)}{-1+\exp(i k_x)}}
  \end{array}
  \right],
\end{align}
from which it is seen that the states on sites $a$ and $c$ (as well as $e$ and $g$) are mutually complex conjugate.
In Fig. \ref{fig3}(c), the DOS is presented, corresponding to the band structure in Fig. \ref{fig3}(b). In the DOS diagram, there are three sharp peaks at the eigenvalues where the three flat bands are located.

\begin{figure*}[htbp]
\centering
  \includegraphics[width=0.7564\textwidth]{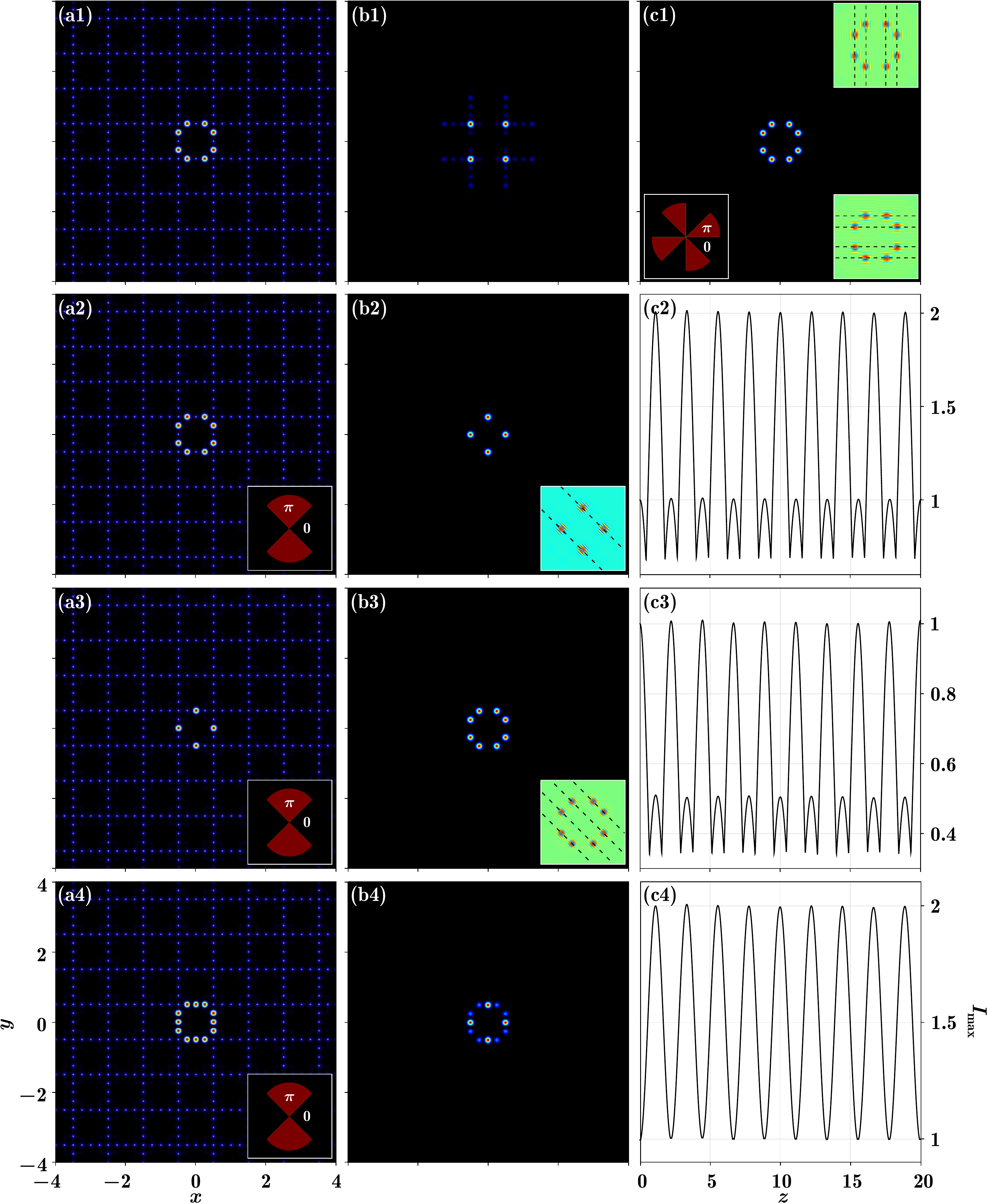}
  \caption{(Color online)
  (a1) Intensity of the input that contains 8 peaks which are exactly located at chosen sites.
  (b1) Discrete diffraction of the input if the peaks are in-phase.
  (c1) Output intensity of the beam if the peaks are out-of-phase.
  Left-bottom inset: phase of the input.
  Right top and bottom insets: interferograms of the output beam with tilted plane waves.
  (a2) Same as (a1), but with a different phase structure.
  Inset: phase of the input beam.
  (b2) Intensity of the beam at $z\approx10$. Inset: interferogram of the output beam with a tilted plane wave.
  (c2) Maximum intensity of the beam versus propagation distance.
  (a3)-(c3) Same as (a2)-(c2), but with only sites $b$ and $f$ excited.
  (a4)-(c4) Same as (a2)-(c2), but for a superposed input that is composed of the inputs in (a2) and (a3).
  All panels except (c2)-(c4) share the same scales and dimensions.
  }
  \label{fig4}
\end{figure*}

Now, we turn to light localization in the Lieb-II lattice, due to flat-band modes.
We first launch a beam with eight peaks (an octupole) into the Lieb-II lattice, such that the peaks excite sites $a$, $c$, $e$ and $g$, as shown in Fig. \ref{fig4}(a1).
If the eight peaks are in-phase, the beam will undergo discrete diffraction during propagation, as exhibited in Fig. \ref{fig4}(b1).
Whereas, if the eight peaks of the beam profile have alternating signs, i.e., they are out-of-phase,
the flat-band mode will be excited, and the beam will not diffract during propagation, as shown in Fig. \ref{fig4}(c1).
There, the left-bottom inset shows the phase distribution of the input beam.
Similar to Fig. \ref{fig2}(c), we also interfere the output beam with tilted plane waves,
and the interferograms are displayed in the right-bottom insets in Fig. \ref{fig4}(c1).
From the interferograms, one can see that the eight peaks are still out-of-phase.
According to the eigenstates listed in Eq. (\ref{eq8}),
one can recognize that the out-of-phase octupole input in Fig. \ref{fig4}(c1) excites the eigenstate $|\beta_4,\mathbf{k}\rangle$,
which belongs to the flat band $\beta_4$.

Next, we still assume that the input beam is an octupole, but the eight peaks are not only not-in-phase, but also not-out-of-phase.
The peaks that excite sites $a$ and $c$ (as well as the peaks that excite sites $e$ and $g$) are in-phase,
but the peaks at sites $a$ ($c$) and $e$ ($g$) are out-of-phase,
as shown in Fig. \ref{fig4}(a2) and the corresponding phase structure in the inset.
One finds that the beam shows neither the discrete diffraction nor the strong localization during propagation,
but it rather exhibits an oscillating behavior.
The energy of the beam moves from sites $a$ and $c$ ($e$ and $g$) to site $b$ ($c$),
to form a quadrupole, as shown in Fig. \ref{fig4}(b2),
and then goes back to the octupole.
This process proceeds periodically and circularly.
If one records the maximum intensity $I_{\rm max}$ of the beam during propagation,
the oscillation can be observed clearly, as presented in Fig. \ref{fig4}(c2),
in which $I_{\rm max}$ is 1 at the initial place and 2 at $z=10$ due to the change from an octupole to a quadrupole.
The interferogram in the inset of Fig. \ref{fig4}(b2) shows that the phase structure is preserved,
even though the intensity structure of the beam varies during propagation.
Since the octupole and the quadrupole mutually transform,
the same oscillating property can be also realized starting from a quadrupole input with alternating signs.
Such a process is displayed in Figs. \ref{fig4}(a3)-\ref{fig4}(c3).
Actually, Figs. \ref{fig4}(a2)-\ref{fig4}(c2) and \ref{fig4}(a3)-\ref{fig4}(c3) incarnate
the reciprocity and the periodicity of this process.

It is interesting to understand why there exists a periodic oscillation during propagation,
and to determine the period.
Since there is no discrete diffraction, the dispersive band modes are not excited.
Considering the flat-band mode corresponding to $\beta_4$, it is excited by the out-of-phase octupole,
hence the inputs in Figs. \ref{fig4}(a2)-\ref{fig4}(a4) can only excite the flat-band modes corresponding to $\beta_{2,6}$.
Therefore, the oscillating property comes from the eigenstates $|\beta_{2,6},\mathbf{k}\rangle$, as displayed in Eq. (\ref{eq8}).
Actually, one can find from the eigenstates $|\beta_{2,6},\mathbf{k}\rangle$ that the sites $a$, $b$, $c$, $e$, $f$ and $g$ are all excited,
and the amplitudes on sites $b$ and $f$ are $\sqrt 2$ times of those on $a$, $c$, $e$ and $g$.
As a result, if the sites $a$, $c$, $e$ and $g$ are excited, sites $b$ and $f$ will be also excited, and \textit{vice versa} ---
this mechanism leads to the periodic oscillation.
Numerical simulations indeed demonstrate that the amplitudes in Figs. \ref{fig4}(b2) and \ref{fig4}(a3) are $\sqrt 2$ times of those in Figs. \ref{fig4}(a2) and \ref{fig4}(b3), respectively.
Concerning the period, one should first determine the energy difference between the two flat bands $\Delta \beta=\beta_2-\beta_6=2\sqrt{2}t$,
and then find the period
\begin{equation}\label{eq9}
D=\frac{2\pi}{\Delta\beta}=\frac{\pi}{\sqrt{2}t}.
\end{equation}
Clearly, the period is associated with the hopping strength $t$, and the bigger the value of $t$, the smaller the period.
In Fig. \ref{fig4}, we set $t=1$, so there are about 9 periods over the distance of 20.
We would like to emphasize that the input can be arbitrarily constructed by the inputs used in Figs. \ref{fig4},
no matter which flat-band mode is excited.
In other words, the freedom to observe the localization due to the flat band is much improved in comparison with earlier investigations.
In Fig. \ref{fig4}(a4), we deliberately designed the input beam as a superposition of the inputs from Figs. \ref{fig4}(a2) and \ref{fig4}(a3).
One finds that the intensity distribution of the corresponding output beam in Fig. \ref{fig4}(b4)
is a superposition of the output intensity distributions from Figs. \ref{fig4}(b2) and \ref{fig4}(b3).
The intensity difference between the octupole and the quadrupole in Fig. \ref{fig4}(b4) is due to the fact that
the maximum intensity in Fig. \ref{fig4}(b2) is 4, while that in Fig. \ref{fig4}(b3) is 1.
From this point of view, one could state that the excited flat-band modes are independent, and they will not affect each other during propagation.
Because of the mutual transformation between the octupole and the quadrupole, the maximum intensity is not a simple sum of those in Figs. \ref{fig4}(c2) and \ref{fig4}(c3).
As shown in Fig. \ref{fig4}(c4), the maximum intensity changes like a cosine curve with the propagation distance.

Last but not least, conical diffraction can be observed if light is launched into the site $b$ and excites the modes around the Dirac cone \cite{leykam.pra.86.031805.2012}.
As discussed above, there is no light located on the site $b$ when the flat-band mode is excited,
therefore, one may construct an input, based on which the flat-band modes and the Dirac cone modes are excited simultaneously.
However, in this paper we omit the discussion of this aspect of beam propagation in the new Lieb lattices.

\section{Conclusion}
\label{conclusion}

In summary, we have constructed two new kinds of face-centered photonic square lattices --- the Lieb-I and the Lieb-II lattices,
which possess even and odd number of flat bands.
Similar to previous investigations, the flat-band modes are strongly localized when they are excited during propagation.
On the other hand, different from previous investigations,
such flat-band modes may exhibit oscillating property during propagation, which has never been observed before.
By choosing certain flat-band modes to describe the beam input, complicated image transmission can be achieved.
Our investigation has potential applications in areas where flat bands can be applied,
and in addition provides a new platform for investigating and understanding the phenomena connected with flat bands.

\section*{Acknowledgement}
The work was supported by the National Basic Research Program of China (2012CB921804),
National Natural Science Foundation of China (61308015, 11474228),
Key Scientific and Technological Innovation Team of Shaanxi Province (2014KCT-10),
and Qatar National Research Fund  (NPRP 6-021-1-005).
MRB also acknowledges support by the Al Sraiya Holding Group.


\bibliographystyle{myprx}
\bibliography{my_refs_library}

\end{document}